\documentclass[aps,prd,preprint,superscriptaddress,tightenlines,nofootinbib]{revtex4}

\usepackage{graphicx}
\usepackage{dcolumn}
\usepackage{bm}
\usepackage{ulem}

\begin{document}

\preprint{CLNS 05/1944}       
\preprint{CLEO 05-30}         

\title{Di-electron Widths of the $\Upsilon(1S,\,2S,\,3S)$ Resonances}


\author{J.~L.~Rosner}
\affiliation{Enrico Fermi Institute, University of
Chicago, Chicago, Illinois 60637}
\author{N.~E.~Adam}
\author{J.~P.~Alexander}
\author{K.~Berkelman}
\author{D.~G.~Cassel}
\author{J.~E.~Duboscq}
\author{K.~M.~Ecklund}
\author{R.~Ehrlich}
\author{L.~Fields}
\author{R.~S.~Galik}
\author{L.~Gibbons}
\author{R.~Gray}
\author{S.~W.~Gray}
\author{D.~L.~Hartill}
\author{B.~K.~Heltsley}
\author{D.~Hertz}
\author{C.~D.~Jones}
\author{J.~Kandaswamy}
\author{D.~L.~Kreinick}
\author{V.~E.~Kuznetsov}
\author{H.~Mahlke-Kr\"uger}
\author{T.~O.~Meyer}
\author{P.~U.~E.~Onyisi}
\author{J.~R.~Patterson}
\author{D.~Peterson}
\author{E.~A.~Phillips}
\author{J.~Pivarski}
\author{D.~Riley}
\author{A.~Ryd}
\author{A.~J.~Sadoff}
\author{H.~Schwarthoff}
\author{X.~Shi}
\author{S.~Stroiney}
\author{W.~M.~Sun}
\author{T.~Wilksen}
\author{M.~Weinberger}
\affiliation{Cornell University, Ithaca, New York 14853}
% ...
\collaboration{CLEO Collaboration} 
\noaffiliation


\date{December 21, 2005}

\newcommand{\gee}{$\Gamma_{ee}$}
\newcommand{\ups}{$\Upsilon$}
\newcommand{\us}{$\Upsilon(1S)$}
\newcommand{\uss}{$\Upsilon(2S)$}
\newcommand{\usss}{$\Upsilon(3S)$}
\newcommand{\ee}{$e^+e^-$}
\newcommand{\mm}{$\mu^+\mu^-$}
\newcommand{\tautau}{$\tau^+\tau^-$}
\newcommand{\ellell}{$\ell^+\ell^-$}
\newcommand{\pipi}{$\pi^+\pi^-$}
\newcommand{\PM}{$\pm$}
\newcommand{\inv}{$^{-1}$}
\newcommand{\bmm}{${\mathcal B}_{\mu\mu}$}
\newcommand{\btt}{${\mathcal B}_{\tau\tau}$}
\newcommand{\geehadtot}{\Gamma_{ee}\Gamma_{\mbox{\scriptsize had}}/\Gamma_{\mbox{\scriptsize tot}}}
\newcommand{\pvis}{P_{\mbox{\scriptsize vis}}}
\newcommand{\ppass}{P_{\mbox{\scriptsize pass given vis}}}
\newcommand{\ehtrig}{\epsilon_{\mbox{\scriptsize htrig}}}
\newcommand{\ecuts}{\epsilon_{\mbox{\scriptsize cuts}}}
\newcommand{\chired}{\chi^2_{\mbox{\scriptsize red}}}

\begin{abstract} 
We determine the di-electron widths of the \us, \uss, and \usss\
resonances with better than 2\% precision by integrating the
cross-section of $e^+e^- \to \Upsilon$ over the \ee\ center-of-mass
energy.  Using \ee\ energy scans of the \ups\ resonances at the
Cornell Electron Storage Ring and measuring \ups\ production with the
CLEO detector, we find di-electron widths of
1.354 \PM\ 0.004 ($\sigma_{\mbox{\scriptsize stat}}$) \PM\ 0.020 ($\sigma_{\mbox{\scriptsize syst}}$) keV,
0.619 \PM\ 0.004 \PM\ 0.010 keV, and
0.446 \PM\ 0.004 \PM\ 0.007 keV for the \us, \uss, and \usss,
respectively.
\end{abstract}

\maketitle

The widths of the \ups\ mesons, $b\bar{b}$ bound states discovered in
1977 \cite{discovery}, are related to the quark-antiquark spatial wave
function at the origin \cite{wavefunction}.  These widths provide a
testing ground for QCD lattice gauge theory calculations
\cite{lattice}.  Improvements in the lattice calculations, such as
avoidance of the quenched approximation \cite{unquenched}, provide an
incentive for more accurate experimental tests.  The di-electron
widths (\gee) of the \us, \uss, and \usss\ have previously been
measured with precisions of 2.2\%, 4.2\%, and 9.4\%, respectively
\cite{pdg}.  Validation of the lattice calculations at an accuracy of
a few percent will increase confidence in similar calculations used to
extract important weak-interaction parameters from data.  In
particular, \gee\ and $f_D$ \cite{fd} provide complementary tests of
the calculation of $f_B$, which is used to determine the CKM parameter
$V_{td}$.

Our measurement of \gee\ follows the method of \cite{pdg}: we
integrate the production cross section of \ups\ over incident \ee\
energies.  If we ignore initial state radiation for clarity, the
partial width is given by
\begin{equation}
\label{eqn:gee}
\Gamma_{ee} = \frac{{M_\Upsilon}^2}{6\pi^2}\int \sigma(e^+e^- \to
\Upsilon) \, dE \mbox{.}
\end{equation}
We also determine the \ups\ full widths using $\Gamma =
\Gamma_{ee}/{\mathcal B}_{\ell\ell}$, where ${\mathcal B}_{\ell\ell}$
is the \ups\ branching fraction to a pair of leptons.

The Cornell Electron Storage Ring (CESR), an \ee\ collider, scanned
center-of-mass energies in the vicinity of the \us, \uss, and \usss,
and the CLEO~III detector collected the \ups\ decay products to
determine the cross section at each energy.  A fit to this resonance
lineshape yields $\int \sigma(e^+e^- \to \Upsilon) \, dE$.  This fit
includes the effects of initial-state radiation, beam energy spread,
backgrounds, and interference between \ups\ and continuum decays.  The
eleven \us\ scans, six \uss\ scans, and seven \usss\ scans have
integrated luminosities of 0.27, 0.08, and 0.22~fb\inv, respectively,
with 0.19, 0.41, and 0.14~fb\inv\ of data below each peak to constrain
backgrounds.

The CLEO~III detector is a nearly 4$\pi$ tracking volume surrounded by
a CsI crystal calorimeter \cite{cleoiii} \cite{driii}.  Charged tracks
are reconstructed in a 47-layer wire drift chamber and 4-layer silicon
strip detector, and their momenta are inferred from their radii of
curvature in a 1.5 T magnetic field.  The calorimeter forms a
cylindrical barrel around the tracking volume, reaching angles
$\theta$ with respect to the beam axis of $|\cos\theta|$ $<$ 0.85,
with endcaps extending this range to $|\cos\theta|$ $<$ 0.98.
Electron showers have a resolution of 75~MeV at 5~GeV (the beam
energy).

The \ups\ mesons are produced nearly at rest and decay into leptonic
final states $e^+e^-$, $\mu^+\mu^-$, or $\tau^+\tau^-$, or into
hadrons via $ggg$, $gg\gamma$, or $q\bar{q}$ intermediate states.  The
\uss\ and \usss\ can also make transitions into other $b\bar{b}$
resonances such as $\chi_{bJ}(nP)$, \us, and \uss.  The leptonic
decays together account for only about 7\% of the decays of each
resonance and are difficult to distinguish from background, so we
select hadrons, fit the hadronic cross section, and report
$\geehadtot$.  We then correct for the missing leptonic modes to
report \gee, assuming ${\mathcal B}_{ee} = {\mathcal B}_{\mu\mu} =
{\mathcal B}_{\tau\tau}$ and obtaining the well-measured \bmm\ from
\cite{istvan}.  (The $\tau$ mass shifts \btt\ below the ${\mathcal
B}_{ee}$ or ${\mathcal B}_{\mu\mu}$ expectation by only 0.05\% at
these energies.)  Thus, $\Gamma_{ee} = (\geehadtot)/(1 - 3{\mathcal
B}_{\mu\mu})$.

\begin{figure}
  \includegraphics[width=0.5\linewidth]{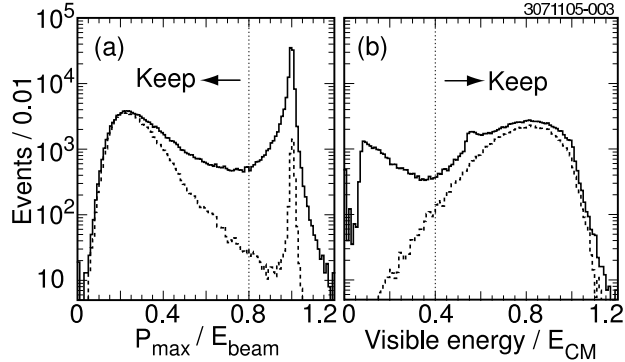}
  \caption{\label{fig:cuts} Two of the distributions used to identify
hadronic \ups\ decays.  Solid histograms are data, dashed are
simulated \us\ decays, both with all hadronic selection criteria
applied except the one shown.  Dotted lines indicate selection
thresholds.}
\end{figure}

Bhabha scattering ($e^+e^- \to e^+e^-$) is our largest potential
background.  We suppress these events by requiring the greatest track
momentum ($P_{\mbox{\scriptsize max}}$) to be less than 80\% of the
beam energy, shown in Figure~\ref{fig:cuts}(a), which reduces the
Bhabha background to approximately the same magnitude as the hadronic
continuum ($e^+e^- \to q\bar{q}$) background.  Continuum annihilation
processes such as these are accounted for by including a $1/s$ term in
the lineshape fit, where $s = {E_{\mbox{\scriptsize CM}}}^2 = (2
E_{\mbox{\scriptsize beam}})^2$.

The contribution of two-photon events ($e^+e^- \to e^+e^- X$) grows
with $\log s$.  We suppress these by requiring the total visible
energy (energy sum of all charged tracks and neutral showers) to be
more than 40\% of the center-of-mass energy, shown in
Figure~\ref{fig:cuts}(b).  The \uss\ and \usss\ additionally have
backgrounds from radiative returns to each lower-energy resonance,
with a cross-section inversely proportional to the initial-state
photon energy.  We therefore add to the fit function a small $\log s$
term (8\% of continuum at 9~GeV) and $1/(\sqrt{s}-M_\Upsilon)$ terms
for \us\ and \uss\ (about 0.5\% of continuum at the \usss).  Because
the off-resonance data are only 20~MeV below each peak, the different
functional forms affect the background estimation at the peak by less
than 0.04\%.

\begin{figure}
  \includegraphics[width=0.5\linewidth]{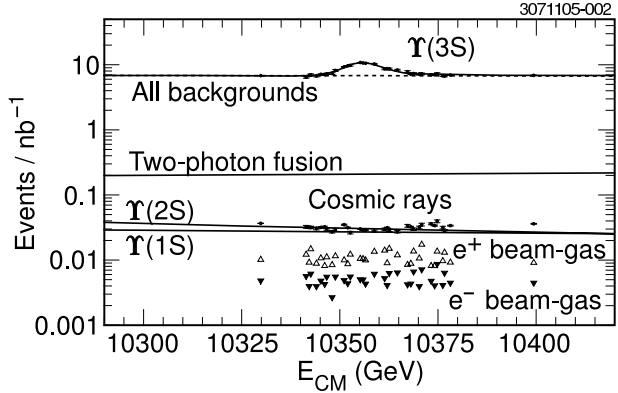}
  \caption{\label{fig:backgrounds} The event yield as a function of
center-of-mass energy in the region of the \usss.  The top points are
data, with the fit superimposed, and the dashed curve represents the
sum of all backgrounds.  The lower points and lines show the
individual non-1/s background contributions.}
\end{figure}

Cosmic rays and beam-gas interactions (collisions between a beam
electron and a gas nucleus inside the beampipe) are suppressed by
requiring charged tracks to point toward the beam-beam intersection
point.  We reduce this to less than 1\% of the continuum by demanding
that at least one reconstructed track pass within 5~mm of the beam
axis and the vertex reconstructed from all primary tracks be within
7.5~cm of the intersection point along the beam axis.  We determine
and subtract the remaining contamination at each energy using special
single-beam and no-beam data runs normalized using events with a
solitary large impact parameter track (for cosmic rays) or vertices
along the beam axis but far from the collision point (for beam-gas).
Individual backgrounds for the \usss\ are illustrated in Figure~\ref{fig:backgrounds}.

While our hadronic selection criteria eliminate essentially all
$\Upsilon \to e^+e^-$ and $\Upsilon \to \mu^+\mu^-$ decays, they
accept 57\% of $\Upsilon \to \tau^+\tau^-$, according to a GEANT-based
Monte Carlo simulation \cite{mc} including final-state radiation
\cite{photos}.  We therefore add to the fit function an $\Upsilon
\to \tau^+\tau^-$ background term, including interference with
continuum $e^+e^- \to \tau^+\tau^-$, using the measured ${\mathcal
B}_{\tau\tau}$ \cite{jean}.

A small fraction of hadronic \ups\ decays fail our event selection
criteria.  Instead of estimating this inefficiency with the Monte
Carlo simulation, which would introduce dependence on the decay model,
hadronization model, and detector simulation, we use a data-based
approach.  We select $\Upsilon(2S) \to \pi^+\pi^- \Upsilon(1S)$ events
to study \us\ decays tagged by \pipi.  If the \pipi\ were sufficient
to satisfy the trigger, the efficiency would be the ratio of \us\
events satisfying our selection criteria (excluding the \pipi\ tracks
and showers) to all \us\ events.

Although this procedure could be applied directly to the \uss\ sample,
the loose two-track trigger involved is prescaled, and thus can only
determine the hadronic efficiency to within 3\% of itself.  Instead,
we use the two-track trigger to determine the efficiency of an
non-prescaled but more restrictive hadronic trigger ($\ehtrig$), and
then use the full statistics from the hadronic trigger to determine
our selection efficiency once this trigger has been satisfied
($\ecuts$).  Our combined event selection and trigger efficiency is
then the product of $\ehtrig$ and $\ecuts$.

\begin{figure}
  \includegraphics[width=0.5\linewidth]{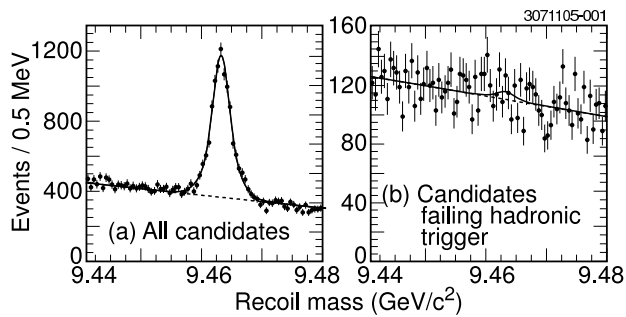}
  \caption{\label{fig:cascades} Mass of the system recoiling against
the \pipi\ in $\Upsilon(2S) \to \pi^+\pi^- \Upsilon(1S)$ candidates
satisfying the two-track trigger, for (a) all events, and (b) events
that satisfy the hadronic trigger.  The dashed curve represents
backgrounds and the solid curve represents the sum of backgrounds and
the recoiling \us\ signal.}
\end{figure}

The mass of the system recoiling against the \pipi\ candidates in the
two-track trigger sample is shown in Figure~\ref{fig:cascades}.  After
correcting for leptonic decays in the \us\ sample, we find $\ehtrig$ =
(99.59 $^{+0.29}_{-0.45}$)\% from the ratio of fit yields.

From $\Upsilon(2S) \to \pi^+\pi^- \Upsilon(1S)$ events that satisfy
the hadronic trigger, we find $\ecuts$ = (98.33 \PM\ 0.33)\%.  This
has been corrected for leptonic decays, the boost of the \us, track
and shower confusion from the \pipi, and the efficiency of the full
set of triggers.  Only the first correction is significant.  Our event
selection and trigger efficiency is therefore (97.93
$^{+0.44}_{-0.56}$)\% for the sum of all non-leptonic \us\ decays.

To find the \uss\ and \usss\ efficiencies, we correct the \us\
efficiency for energy dependence and for transitions specific to these
excited states, using simulations.  Energy dependence is negligible;
only transitions to lower \ups\ resonances which then decay to \ee\ or
\mm\ introduce a significant loss of efficiency.  We measure the
branching fractions of these decays to be (1.58 \PM\ 0.16)\% and (1.34
\PM\ 0.13)\%, respectively, resulting in \uss\ and \usss\ efficiencies
of (96.18 $^{+0.44}_{-0.56}$ \PM\ 0.15)\% and (96.41
$^{+0.44}_{-0.56}$ \PM\ 0.13)\%.  Both uncertainties are statistical,
but the first is common to all three resonances.

We use Bhabha events to determine the relative luminosity of each scan
point.  We select the Bhabhas by requiring two or more central tracks
with momenta between 50\% and 110\% of the beam energy, and a ratio of
shower energy to track momentum consistent with $e^+$ and $e^-$.
Contamination from $\Upsilon \to e^+e^-$ is 2--5\% and is readily
calculated given ${\mathcal B}_{ee}$ once we have done our \ups\
lineshape fit.  Our subtraction includes energy-dependent interference
between $\Upsilon \to e^+e^-$ and Bhabhas.

We determine the overall luminosity scale using the method of
\cite{oldlumi} from Bhabhas, $e^+e^- \to \mu^+\mu^-$, and $e^+e^- \to
\gamma\gamma$, with the {\textsc Babayaga} event generator
\cite{babayaga}.  The systematic uncertainties from the three
processes are 1.6\%, 1.6\%, and 1.8\%, respectively, dominated by
track finding and resonance interference for \ee\ and \mm, and by
photon finding and angular resolution for $\gamma\gamma$.  The three
measurements give consistent results off-resonance, where \ups\
contamination is negligible.  We use the weighted mean to determine
the luminosity, and take the RMS scatter of 1.3\% as the systematic
uncertainty.

Bhabha and $\gamma\gamma$ luminosities, normalized to the same value
off-resonance, deviate by (0.8 \PM\ 0.2)\%, (0.3 \PM\ 0.4)\%, and (0.7
\PM\ 0.2)\% at the \us, \uss, and \usss\ peaks.  We correct each \gee\
by half of its discrepancy and take half the discrepancy and its
uncertainty in quadrature as a systematic uncertainty.

Accurate measurement of beam energies are also needed to determine
\gee.  An NMR probe calibrates the field of the CESR dipole magnets
and hence provides the beam energy, after corrections for RF frequency
shifts, steering and focusing magnets, and electrostatic
electron-positron separators.  To limit our sensitivity to drifts in
this measurement, we limit scans to 48 hours and alternate
measurements above and below the peak.  By repeating a resonance
cross-section measurement at a point of high slope, we find that the
beam energy calibration drifts by less than 0.04~MeV within a scan (at
68\% confidence level), which implies a 0.2\% uncertainty in \gee.

The data for each resonance are separately fit to a lineshape that
consists of a three-fold convolution of (a) a Breit-Wigner resonance
including interference between $\Upsilon \to q\bar{q}$ and $e^+e^- \to
q\bar{q}$ with zero phase difference at $\sqrt{s} \ll M_\Upsilon$, (b)
an initial-state radiation distribution as given in Equation (28) of
\cite{kf}, and (c) the Gaussian spread in CESR beam energy of about
4~MeV, plus the background terms described above.  The radiative
corrections account for emission of real and virtual photons by the
initial \ee.  We do not correct for vacuum polarization, which is
absorbed into the definition of \gee.  The resulting \gee\ therefore
represents the Born diagram coupling of a pure \ee\ state to the \ups.
The lineshape fits are insensitive to the Breit-Wigner widths at the
0.1\% level, so we fix these widths to their PDG values \cite{pdg}.
The value of $\geehadtot$ of each resonance is allowed to float, as is
the continuum normalization, and, to remove sensitivity to beam energy
shifts between scans, the peak energy of each scan.  In addition, we
fit for the beam energy spread of groups of scans with common CESR
horizontal steerings, but allow shifts when the steerings change,
since they can change the beam energy spread by 1\%.

\begin{figure*}
  \includegraphics[width=\linewidth]{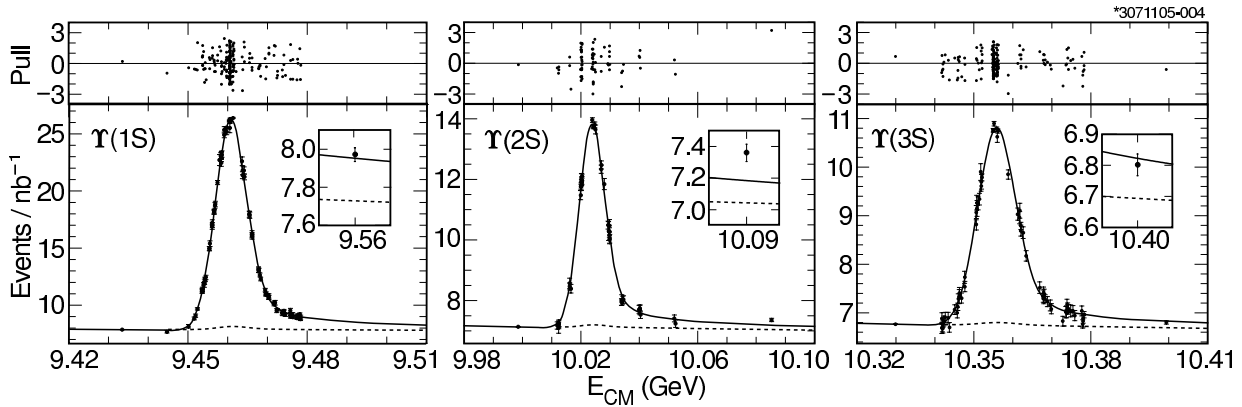}
  \caption{\label{fig:fits} The hadronic yield versus center-of-mass
energy in the vicinity of the three \ups\ resonances.  Points
represent the data, corrected for fitted beam energy shifts between
scans, the solid line is the fit, the dashed line is the sum of all
backgrounds, and the insets show high-energy measurements.  The pull
of each point is shown above.}
\end{figure*}

The fit results are plotted in Figure~\ref{fig:fits}.  The fit
function describes the data well, though it results in larger $\chi^2$
values for the \us\ and \uss.  The $\chi^2$ per degree of freedom
($N_{\mbox{\scriptsize dof}}$) for \us\ is $240/187$ (0.5\% confidence
level), for \uss\ is $107/66$ (0.1\% confidence level), and for \usss\
is $155/159$ (59\% confidence level).  We see no obvious trends in
pull (residual divided by uncertainty) versus energy or versus date,
so we take the large $\chi^2$ values as an indication that
point-to-point uncertainties are underestimated, and add
$\sigma_{\mbox{\scriptsize stat}}\sqrt{\chi^2/N_{\mbox{\scriptsize
dof}} - 1}$ to the systematic uncertainty, if
$\chi^2/N_{\mbox{\scriptsize dof}} > 1$.  This effectively multiplies
the statistical uncertainty ($\sigma_{\mbox{\scriptsize stat}}$) by
$\sqrt{\chi^2/N_{\mbox{\scriptsize dof}}}$.  All uncertainties are
listed in Table~\ref{tab:unc}.

\begin{table}
  \caption{\label{tab:unc} All uncertainties in \gee.  The correction
  for leptonic modes is made for \gee\ but not $\geehadtot$.  The
  uncertainties in hadronic efficiency and overall luminosity scale
  are common to all three resonances.}
  \renewcommand{\arraystretch}{1.25}
  \begin{tabular}{l c c c}
    \hline\hline Contribution to \gee & \hspace{0 cm}\us\hspace{0 cm} & \hspace{0 cm}\uss\hspace{0 cm} & \hspace{0 cm}\usss\hspace{0 cm} \\\hline
    Correction for leptonic modes        	   & 0.2\%  & 0.2\%  & 0.3\%  \\
    Hadronic efficiency                            & 0.5\%  & 0.5\%  & 0.5\%  \\
    $Xe^+e^-$, $X\mu^+\mu^-$ correction  	   & 0      & 0.15\% & 0.13\% \\
    Overall luminosity scale                       & 1.3\%  & 1.3\%  & 1.3\%  \\
    Bhabha/$\gamma\gamma$ inconsistency  	   & 0.4\%  & 0.4\%  & 0.4\%  \\
    Beam energy measurement drift \hspace{0.5 cm}  & 0.2\%  & 0.2\%  & 0.2\%  \\
    Fit function shape                   	   & 0.1\%  & 0.1\%  & 0.1\%  \\
    $\chi^2$ inconsistency               	   & 0.2\%  & 0.6\%  & 0      \\\hline
    Total systematic uncertainty         	   & 1.5\%  & 1.6\%  & 1.5\%  \\
    Statistical uncertainty              	   & 0.3\%  & 0.7\%  & 1.0\%  \\\hline
    Total                                	   & 1.5\%  & 1.8\%  & 1.8\%  \\\hline\hline
  \end{tabular}
\end{table}

We assume that $e^+e^- \to q\bar{q}$ interferes only with the
$q\bar{q}$ component of hadronic \ups\ decays.  The \us\ fit favors
this interference scheme over the no-interference hypothesis by 3.7
standard deviations.  It is also possible that $e^+e^- \to q\bar{q}
\to$ hadrons interferes with $\Upsilon \to ggg \to$ hadrons.  If so,
full interference between all final states, all with a common phase
difference near $\pm\pi/2$ ($\Upsilon \to ggg$ \mbox{phase}
\mbox{minus} $\Upsilon \to q\bar{q}$ phase), would shift $\geehadtot$
by $\mp$5.4\%, $\mp$3.8\%, and $\mp$3.5\% for the \us, \uss, and
\usss, respectively \cite{thesis}.  This is the most extreme case.
Overlap of isospin and flavor states for these two processes suggest
that this interference, if it occurs, affects $\geehadtot$ at no more
than the $\sim$1\% level.

Our values of $\geehadtot$, listed in Table \ref{tab:results}, are
consistent with, but more precise than, the PDG world averages
\cite{pdg} and our \usss\ measurement is substantially more precise.
Also listed in the Table are the di-electron widths and ratios of
these widths, in which common systematic uncertainties have been
canceled.  Assuming ${\mathcal B}_{ee} = {\mathcal B}_{\mu\mu}$
and using \cite{istvan}, we obtain new values of the \ups\ full
widths: 54.4 \PM\ 0.2 ($\sigma_{\mbox{\scriptsize stat}}$) \PM\ 0.8
($\sigma_{\mbox{\scriptsize syst}}$) \PM\ 1.6 ($\sigma_{{\mathcal
B}_{\mu\mu}}$) keV for the \us, 30.5 \PM\ 0.2 \PM\ 0.5 \PM\ 1.3 keV
for the \uss, and 18.6 \PM\ 0.2 \PM\ 0.3 \PM\ 0.9 keV for the \usss.

\begin{table}
  \caption{\label{tab:results} The results of $\geehadtot$ for the
three resonances, the di-electron widths \gee, and their ratios.
The first uncertainty is scaled statistical and the second
is systematic.}
  \renewcommand{\arraystretch}{1.25}
  \begin{tabular}{c c c}
    \hline\hline 
    $\geehadtot(1S)$ & & 1.252 \PM\ 0.004 \PM\ 0.019 keV                   \\
    $\geehadtot(2S)$ & & 0.581 \PM\ 0.004 \PM\ 0.009 keV                   \\
    $\geehadtot(3S)$ & & 0.413 \PM\ 0.004 \PM\ 0.006 keV                   \\\hline
    $\Gamma_{ee}(1S)$ & \hspace{1.59 cm} & 1.354 \PM\ 0.004 \PM\ 0.020 keV \\
    $\Gamma_{ee}(2S)$ & & 0.619 \PM\ 0.004 \PM\ 0.010 keV                  \\
    $\Gamma_{ee}(3S)$ & & 0.446 \PM\ 0.004 \PM\ 0.007 keV                  \\\hline
    $\Gamma_{ee}(2S)/\Gamma_{ee}(1S)$ & & 0.457 \PM\ 0.004 \PM\ 0.004      \\
    $\Gamma_{ee}(3S)/\Gamma_{ee}(1S)$ & & 0.329 \PM\ 0.003 \PM\ 0.003      \\
    $\Gamma_{ee}(3S)/\Gamma_{ee}(2S)$ & & 0.720 \PM\ 0.009 \PM\ 0.007      \\\hline\hline
  \end{tabular}
\end{table}

We gratefully acknowledge the effort of the CESR staff 
in providing us with excellent luminosity and running conditions.
This work was supported by 
the A.P.~Sloan Foundation,
the National Science Foundation,
and the U.S. Department of Energy.

\end{document}